\documentclass[12pt,manuscript]{aastex}
\usepackage{natbib}
\usepackage{amssymb,amsmath}
\newcommand{\Msun}{M_{\rm \odot}}
\newcommand{\Rsun}{R_{\rm \odot}}
\newcommand{\Lsun}{L_{\rm \odot}}
\newcommand{\Teff}{T_{\rm eff}}
\newcommand{\loggteff}{\log g- \log T_{\rm eff}}

\def\note #1]{{\bf #1]}}
\def\fig{.}

\slugcomment{{\sl The Astrophysical Journal}; Accepted}

\shorttitle{Objective Stellar Parameters}
\shortauthors{Quirion et al.}

\begin{document}

\title{Automatic Determination of Stellar Parameters via Asteroseismology of Stochastically Oscillating Stars: Comparison with Direct Measurements} 

\author{Pierre-Olivier Quirion$^{1,2,3}$, and J\o rgen Christensen-Dalsgaard$^{2,3}$, Torben Arentoft$^{2,3}$}
\affil{$^1$ 
Canadian Space Agency, 6767 Route de l'A\'eroport, Saint-Hubert, QC, J3Y 8Y9 Canada 
\\
$^2$ Department of Physics and Astronomy, Aarhus University,
DK-8000 Aarhus C, Denmark
\\
$^3$ Danish AsteroSeismology Center
}

\email{pierre-olivier.quirion@asc-csa.gc.ca,jcd@phys.au.dk,ta@vucaarhus.dk}

\begin{abstract}
Space-based projects are providing a wealth of high-quality asteroseismic data, including frequencies for a large number of stars showing solar-like oscillations. These data open the prospect for precise determinations of key stellar parameters, of particular value to the study of extra-solar planetary systems. Given the quantity of the available and expected data it is important to develop efficient and reliable techniques for analyzing them, including the determination of stellar parameters from the observed frequencies. Here we present the SEEK package developed for the analysis of asteroseismic data from the {\it Kepler} mission. A central goal of the package is to obtain a fast and automatic determination of the stellar radius and other parameters, in a form that is statistically well-defined. The algorithms are tested by comparing the results of the analysis with independent measurements of stellar radius and mass, for a sample of well-observed stars. We conclude that the SEEK package fixes stellar parameters with accuracy and precision.
\end{abstract}


\section{ASTROPHYSICAL CONTEXT}

Solar-like oscillation occurs in stars with convective envelopes and are seen as acoustic modes (p modes) with low angular degree ${l}\lesssim 3$ and intermediate to high radial order $10\lesssim n \lesssim 30$. After early work by \citet{AndOsa75} and \citet{GolKee77a,GolKee77b}, evidence has been found that the excitation of solar-like oscillation is caused by turbulent convection in the uppermost part of the envelopes of stars \citep[see for example][]{Balmfo1992,GooGouKos92,GolMurKum94}. An important aspect of the driving caused by turbulent convection is that the excitation of the mode occurs at random times, and hence the process is known as stochastic excitation. One of the effects of this stochastic behavior is that the eigenmode phases are changing with time. In the Sun, the modes with the largest amplitudes, around 3000 $\mu$Hz, have a coherence time of less than 12 days \citep{Gel02,Cha97b}. We can contrast this value to classical pulsators where the coherence time
  is of the order of the stellar evolution time scale. Short but recent introductions to the topic of stochastic modes excitation, and references to modern approaches to that problem, can be found in \citet{Houdek2006} and \citet{Sam08}.

The first star other than the Sun showing evidence for individual frequencies of solar-like oscillation, $\eta$ Bo\"otis, was observed by \citet{Kje95}, using a technique measuring the equivalent widths of the stellar Balmer lines over time. Most subsequent observations of solar-like oscillation have been made with Doppler-velocity measurements using a laboratory reference (typically an iodine cell) to measure the shift in a stellar spectrum. This technique finally permitted the first unambiguous detection of solar-like oscillation in $\alpha$~Centauri A \citep{BouCar01,BouCar02}. Since then, p-mode excitation has been detected with ground-based observations in more than a dozen main-sequence and subgiant stars. Lists of these solar-like oscillators can be found in \citet{BedKje08} and \citet{Bru10}.

Doppler-velocity measurement is the technique of choice to observe solar-like oscillation, providing the least interference from stellar background contributions to the `noise' in the observations. The ratio between the signal and the stellar background noise, that is non-periodic signals coming, for instance, from granulation, is far higher in Doppler-velocity than in photometric intensity measurements.  Solar-like oscillations can also be observed in photometric intensity measurement, albeit with a higher stellar background noise, but with the advantage of allowing simultaneous observation of a large number of stars. However, owing to noise induced by the Earth's atmosphere such observations are essentially only possible from space \citep{Harvey88}. Space-based photometric data for solar-like oscillations have started to be obtained with the help of the CoRoT \citep{Baglin2006} and {\it Kepler} \citep{Boruck2010} missions. Results coming from these satellites, in particular
  the {\it Kepler} satellite \citep{Gillil2010}, are indeed a motivation for the work undertaken in the present study. The development of an objective technique to interpret results from asteroseismology, more than being intrinsically desirable, is now made necessary by the era of high-speed space photometry, where data bases containing large numbers of light curves ready for asteroseismic study are made available. By objective asteroseismology we mean a technique that is not directed at every step by subjective choice of a scientist and that results in an estimate of global stellar parameters along with realistic, if not completely objective, uncertainties on these parameters. The need for efficient and reliable, automatic analysis of asteroseismic data will be further emphasized with the ESA PLATO mission \citep{Catala2009} which, if finally selected, will be launched around 2018.

The Sun is evidently the canonical example of solar-like oscillation. As shown in detail in the review on helioseismology by \citet{Christensen-Dalsgaard02}  the study of the Sun's oscillations has proven that physical information can be extracted from the understanding of its vibration modes. The extraction of information from other solar-like stars has started to occur and the sustained modeling efforts by several groups has yielded many asteroseismic studies  \citep[e.g.][]{Tho03,Egg04,MigMon05,CarEggBou05,EggCarBou05,Egg08,Mos08b}. This has positively confirmed the promise of asteroseismology: to dramatically improve knowledge about stellar parameters. This prospect, especially the promise to determine stellar radii within a 3\,\% error, is very attractive for the exoplanet groups involved in the {\it Kepler} mission. For example, \citet{Bai07} showed how the interferometric radius measurement of HD~189733 has dramatically improved and clarified the value for the radius a
 nd mean density of its planetary companion. We expect radius measurement from asteroseismology to have similar effect. An early indication of the power of {\it Kepler} asteroseismology to characterize exoplanet hosts was provided by \citet{Christ2010}.

Photometric measurement of planetary transits, given that a precise modeling of the limb darkening is achieved (Prsa et al. 2010, in preparation), permits to identify four parameters of the star-planet systems: its orbital period $\Pi_{\rm orb}$, its inclination $i$ toward the observer and the fractional radii of the star and planet, which are defined as 
\begin{eqnarray}\label{radius}
r_{\rm A}=\frac{R_{\rm A}}{a}{\rm,} &{\rm and}& r_{\rm b}=\frac{R_{\rm b}}{a}
\; ,
\end{eqnarray} 
respectively. We clearly see that the knowledge of the stellar host radius $R_{\rm A}$ and its associated error directly leads to a determination of the desired planet radius $R_{\rm b}$, as well as the orbital semi-major axis $a$. Further details on the influence of the precision of stellar parameters on the inference of physical properties of planetary system were provided by \citet{Southworth09}. One of the goals of the present study is to test the reliability of asteroseismology in determining the radius of stars by comparing the results with available interferometric measurements.

This work presents an objective procedure to extract information from solar-like oscillations. The procedure, named SEEK, determines fundamental stellar parameters with the use of quantities inspired by the p-mode asymptotic asteroseismic relations, namely the large and small separations, $\Delta \nu$ and $\delta \nu$, respectively \citep{Tassoul80}. It can also make use of standard astrophysical input, the parallax $\pi$, the $V$ band magnitude and the reddening $E_{B-V}$, as well as input derived from atmosphere modeling, the effective temperature $\Teff$, the gravity $\log g$ and finally the iron to hydrogen ratio [Fe/H]. Therefore, SEEK is not a purely asteroseismic procedure that only uses the frequency spectra of a pulsating star to constrain its fundamental parameters, but rather a hybrid procedure that uses both traditional and seismic input.

The procedure is part of the Kepler Asteroseismic Science Consortium (KASC) pipeline that provides in priority stellar parameters to be used by exoplanet seekers of the {\it Kepler} team. Its aims are also to analyze every star in which $\Delta \nu$ and/or $\delta \nu$ has been fixed and to provide an extended and homogeneous overview of the disk population observed by {\it Kepler}. A description of SEEK and of its main characteristics is made in the following sections. We tested SEEK on a set of well-studied stars. Of course the Sun is the golden standard, but we also tested the procedure against stars with diameters determined through interferometric measurements and/or with masses derived from Kepler's third law, thus testing our procedure against the closest we have to absolute quantities. The stars studied comprise $\beta$~Hydri (HD~2151), $\tau$~Ceti (HD 10700), Procyon~A (HD~61421), $\eta$~Bo\"otis (HD~121370), $\alpha$~Centauri~A and B (HD~128620 and HD~128621), and $
 70$~Ophiuchi A (HD~165341).

\section{THE SEEK PROCEDURE}

\subsection{Stellar models}
The SEEK procedure makes use of a large grid of stellar models computed with the Aarhus Stellar Evolution Code (ASTEC). It compares an observed star with every model of the grid and makes a probabilistic assessment, with the help of Bayesian statistics, about the structure of that star.

Details of ASTEC and its input physics were described by \citet{Christensen-Dalsgaard08a}. Out of the possible options offered by ASTEC, we used, for all our computations: the OPAL equation of state \citep{Swenso1996} along with the OPAL plus Ferguson \& Alexander opacity tables \citep{IglRog96,AleFer94}, the element to element ratios in the metallic mixture of \citet{GreSau98}, and convection treated with the mixing-length formulation of \citet{Boehm-Vitense58}; the mixing length to pressure scale height ratio $\alpha$, characterizing the convective efficacy, was treated as a variable parameter in the SEEK fits. Neither diffusion nor overshooting was included. 

A brief introduction to a beta version of SEEK was presented along with a series of tests using synthetic observations in $\S$ 5 of \citet{Ste09}. The beta version was built with an ad hoc probabilistic approach that is quite different from the one used in the present version. There is a major difference in the way errors on stellar parameters is obtained. This difference lies in the fact that the new grids have been extended and are dense enough to permit the use of Bayesian statistic. While SEEK yields similar results with both approaches, the Bayesian method used here is more rigorous, more robust and more comprehensive than the technique developed for the beta version of SEEK. 

The core of SEEK is a grid of models used to fit observations.  We have calculated 7,300 evolution tracks or 5,842,619 individual models. Each track calculation was stopped after the track had reached the giant branch or an age of $\tau = 15 \times 10^9$~yr. The tracks are separated into 100 subsets using a different combination of metallicity $Z$, initial hydrogen content $X_0$ and mixing-length parameter $\alpha$. These combinations are separated into two regularly spaced and interlaced subgrids. The first subgrid comprises tracks with $Z = [0.005,0.01,0.015,0.02,0.025,0.03]$, $X_0 = [0.68,0.70,0.72,0.74]$, and $\alpha=[0.8,1.8,2.8]$ while the second subset has $Z = [0.0075,0.0125,0.0175,0.0225,0.0275]$, $X_0 = [0.69,0.71,0.73]$, $\alpha=[1.3,2.3]$. Every subset is composed of 73 tracks spanning from 0.6 to 3.0~$\Msun$. The spacing in mass between the tracks is 0.02 $\Msun$ from 0.6 to 1.8 $\Msun$ and 0.1 from 1.0 to 3.0 $\Msun$. A slice of the grid is presented in Figure \
 ref{fig:slice} for $Z = 0.015$, $X_0 = 0.72$ and $\alpha = 1.8$ in a $\loggteff$ diagram.

The structure of the SEEK procedure is quite flexible. It will be easy in future to expand the grid with regularly or irregularly spaced tracks. In addition, we have started to test the possibility of interpolating models along the age-mass plane of the grid and between grid values of $Z$ and $X_0$. The results are quite promising and it seems that the grid is dense enough with regards to these parameters to realize precise interpolation. However, interpolation tests in the $\alpha$ direction were not convincing and it seems that the grid would have to be refined significantly with respect to that parameter to permit precise interpolation.

Some care has been required in defining the parameters of the grid, including the specific range and spacing we have chosen. Several factors influenced the definition of the composition. Recent analyses have led to substantial revisions of the solar surface composition \citep{Asplun2005, GreAspSau07, Asplun2009}; since the solar composition is used as reference to relate the observed [Fe/H] to $Z/X$ this potentially affects the assumed stellar composition. Here we have chosen to use the older value of \citet{GreSau98}. This can be justified based on the fact that solar models using the old values are closer to the helioseismic inferences than those using the more recent abundances \citep[see, for example][]{Bas07,BasAnt08}. We also emphasize that for most stars, atmospheric parameters in the literature are still only available with older metallic mixtures. Homogeneity of our analysis requires the use of these mixtures. The metallicity range used in the grid, $ -0.61 \lesssim 
 {\rm [Fe/H]} \lesssim 0.20$, covers most ($\gtrsim 90\%$) of the stellar population of the disk if the Hipparcos color-magnitude diagram is taken as a reference \citep{Reid99,Chabrier01}. This should permit the analysis of a large majority of the {\it Kepler} stars.
 
The careful reader will also remark that relatively high values of $Y_{\odot} = 0.2713$ and $Z_{\odot} =0.0196$  have been used in the standard definition of [Fe/H]. This value is used to calibrate solar models of ASTEC to the right luminosity \citep{Christensen-Dalsgaard98}. This value is used in all conversions done here.

The mass step of 0.02 $\Msun$ from 0.6 to 1.8 $\Msun$ has first been chosen to be small enough to permit interpolation in the $\loggteff$ diagram. However, we realized that this small step in mass has the virtue of making the grid sufficiently dense that we do not have to rely on interpolation to make our probabilistic estimate of the stellar parameters. The lower limit to the range results from the fact that molecules start to play an important role in these stars and that our evolution code does not include them. This should not be a handicap to SEEK since the low absolute magnitude of the stars not included in the grid, with $M < 0.5$~$\Msun$, makes them improbable {\it Kepler} targets. The upper limit of 1.8~$\Msun$ is approximately the highest mass of the grid at which a model on the main sequence has at least 1\% of its radius as a convective envelope. We wanted to keep a high density of models up to that limit where it is most likely to observe stars with stochasticall
 y excited oscillations. The models at higher masses are put as a safeguard permitting us to pin down a star with a very thin convective zone showing solar-like oscillations. We also remark that typically models with higher $\alpha$ and/or higher $X_0$ and/or higher $Z$ will tend to keep a convective envelope at slightly higher masses. Similarly, these models develop a convective core, leading to the typical hook path in the H-R diagram at higher masses. 

The grid mid-value of $\alpha =1.8$ was chosen in accordance with the solar calibration. The span of $\alpha$ is a trade-off between trying to get a wide range of values, from 0.8 to 2.8, and to limit the quantity of computed models in the grid. We had a bias towards testing a wide range of convective efficiency since it is not clear how the mixing-length theory extends from the Sun to other main-sequence stars. However, this choice is not ideal since, as mentioned, the steps are too large to permit interpolation between the grid points.

\begin{figure}
\centering
\includegraphics[angle= 0, width=12cm]{\fig/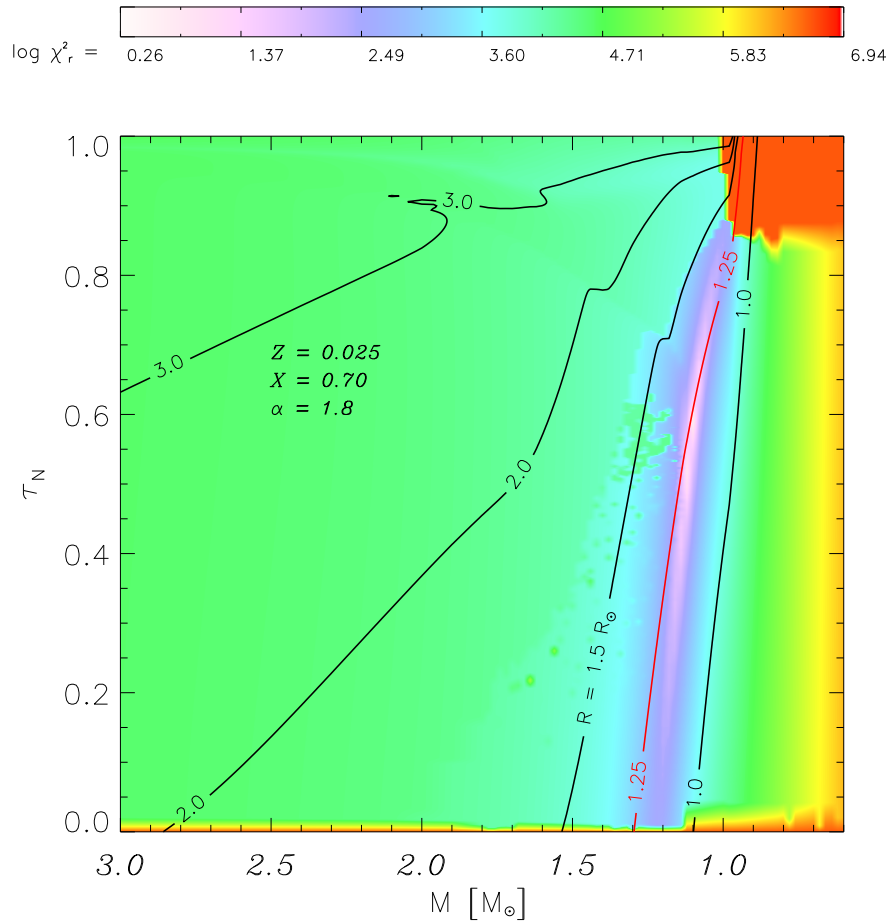}
\caption{
Slice of the solution valley obtained by SEEK for $\alpha$~Cen~A at 
$Z =0.025$, $X_0 = 0.70$, and $\alpha=1.8$. Note the logarithmic color scale showing the value of $\chi^2_r$ as a function of $M$ and $\tau_{\rm N}$
(see Eq. \ref{norm_age}).
The solution is centered on the constant radius line of $R = 1.25 \, \Rsun$, shown as a solid red line. The solid black lines are also radius isocurves. Models older than 13~Gyr, on the upper right region of the figure, and the models on the ZAMS, with $\tau_{\rm N} = 0 $, were excluded from the fit and given values of $\log \chi^2_r = 6.0$. The slight deviation in the constant-radius isocurves with $R \geq 1.5 \, \Rsun$ and $\tau_{\rm N} \gtrsim 0.7$ marks the position of the hook in the evolution tracks.
\label{fig:valley}}
\end{figure}

The role of SEEK is more to measure the size of the ``valley of good solutions'' produced by {\it Kepler} observations than it is to probe the deepest abyss of that valley. Specifically, the quality of a solution is measured by how well the computed
observables in the grid, $\{q_i^{\rm g}\}$, match the observed values
$\{q_i^{\rm obs}\}$; this is quantified in terms of
\begin{equation}\label{chi2}
\chi^2 = {\displaystyle\sum_{i=0}^{n}}\left( \frac{{q}_i^{\rm obs} - {q}_i^{\rm g}}{\sigma_i} \right)^2 \; ,
\end{equation} 
where $\sigma_i$ is the estimated error for the observation $q_i^{\rm obs}$ and $n$ the number of observables. 
SEEK does not claim to find the best model for an observed star; its aim is to draw the contour of good solutions which is located around $\chi^2_r = \chi^2/P \sim 1$, $P$ being the number of degrees of freedom of the problem. SEEK also outputs the stellar parameters with reliable error bars. We do not want to underestimate the errors by restricting the size of our grid. The approach is especially well suited to our problem since our solutions coming from the observation of $\Delta \nu$ are always heavily degenerate. This means that we do not expect a single and small region of the parameter space to yield a result that is much better than any other in terms of $\chi^2$. Instead we have, as illustrated in Figure \ref{fig:valley} for the star $\alpha$~Centauri~A, a large region of the parameter space with $\chi^2_r \sim 1$. In Figure \ref{fig:valley}, we present for drawing convenience $\log \chi^2_r$ as a function of the mass $M$ and the normalized age defined as
\begin{equation}\label{norm_age}
\tau_{\rm N} = \frac{\tau}{\min(\tau_{\rm RGB},14{\rm Gyr})} \; ,
\end{equation} 
where $\tau$ is the age of a model on a specific track and $\tau_{\rm RGB}$ is the age of the model following the same track when it reaches the red-giant branch (RGB). In Figure \ref{fig:valley}, this parameterization is useful for visual purposes as it permits a smooth transition of the $\chi^2$ values from the lower-mass models not reaching the RGB to the models reaching it. The exact position of $\tau_{\rm RGB}$ on the RGB is not critical since the evolution time scale on the RGB is much faster than on the main sequence. We stress that this parameterization was used for cosmetic purposes and is not involved in the SEEK computations. The models showing higher $\chi_r^2$ to the left of the valley are caused by the appearance of convective cores in heavier models; this makes the automatic computation of the small separation somewhat difficult and creates these artifacts.

{
Two works presenting pipelines to do asteroseismology of solar-like stars have recently been published, \citet{Metcal2009} and \citet{Basu2010}. In the former, the authors combine a pipeline using a genetic algorithm to find the best model in the parameter space and a local analysis, based on linearization around this solution, to refine the fit and estimate the error in that solution. \citet{Basu2010} find the best model with a grid of stellar models, not unlike SEEK but using a smaller grid, while the errors are determined with a frequentist approach, using a Monte Carlo simulation of synthetic observations. We certainly recognize the desirability of a genetic algorithm, or a similar technique aiming at the absolute minimum in the parameter space, while also acknowledging the large computation power needed for such techniques.
However, we note that a local error estimator used by \citet{Metcal2009} assumes that the solution is linear within the error bars and that no nearby local minima can be deep enough to contribute to the error analysis;
this can lead to underestimation of the errors. 
In our view, only an error estimator, Bayesian or frequentist, looking at various models around the best solutions may provide fully reliable results. 
This is particularly true in a problem, as the present, where the valley of solution is extended. This preference for Bayesian statistics has motivated our use of it in the development of SEEK. 
}

\subsection{The Observables}

The grid permits the mapping of the model physical input parameters $\boldsymbol{p} \equiv \{M,\tau,Z,X_0,\alpha\}$ into the grid of observable quantities $\boldsymbol{q}^{\rm g} \equiv \{\Delta\nu,\delta\nu,\Teff,\log g,{\rm [Fe/H]},M_V,...\}$, defining the transformation 
\begin{equation}\label{grid}
\boldsymbol{q}^{\rm g} = \mathcal{K}(\boldsymbol{p}).
\end{equation}
We use these quantities and compare them to the actual observed quantities $\boldsymbol{q}^{\rm obs}$. 

Some of the observables used for the subsequent fit are easily extracted from the grid like $\Teff$, $\log g$, [Fe/H]. However, $M_V$, $\Delta \nu$, and $\delta \nu$ need some extra attention. We compute
absolute magnitude $M_V$, in the Johnson $V$ band,
of our model by applying the bolometric correction $B_{\rm c}(V)$ of the \citet{VanCle03} tables to our luminosity $L$,
\begin{equation}\label{magnitude}
M_V = M_{\rm Bol ,\odot} - 2.5 \log \frac{L}{\Lsun} - B_{\rm c}(V),
\end{equation} 
where $M_{\rm Bol , \odot}$ is the \citet{VanCle03} prescription of 4.75 for the Sun's bolometric magnitude. We can then compare this value to its observed counterpart defined as
\begin{equation}\label{parallax}
M^{\rm obs}_V =  5\log \pi-(V-A_v)+5.
\end{equation} 
The conversion is done using the following relation between the absorption $A_V$ and the interstellar reddening $E_{B-V}$: $A_V = 3.1 E_{B-V}$.

We computed p-mode frequencies for all models on the main sequence and subgiant branch using the ADIPLS code \citep{Christensen-Dalsgaard08b}; the models on the red-giant branch are not included in this asteroseismic study. The large separation $\Delta \nu$ is computed with modes of $l =0$ only,
\begin{equation}\label{large}
\Delta \nu = \langle \nu_{n,0} -  \nu_{n-1,0} \rangle \; ,
\end{equation} 
while $\delta \nu$ is the combination of modes with $l = 0$, $2$ only, 
\begin{equation}\label{small}
\delta \nu = \langle \nu_{n,0}  - \nu_{n-1,2} \rangle \; ;
\end{equation} 
here $\nu_{n,l}$ is the frequency of the pulsation mode with angular degree $l$ and radial order $n$. We know that these asymptotic values change with the radial order $n$ in models. This is especially true in the case of $\delta \nu$. To compare observed separations with model values we have computed the average separations with up to 8 modes having different $n$ and centered on a predetermined value of $\nu_{\rm central}$. The quantity $\nu_{\rm central}$ is related to the maximum power $\nu_{\rm max}$ seen in the Fourier transform of light curves of solar-like stars \citep[see][for comments on $\nu_{\rm max}$]{Ste09}. Thus, in our grid the large and small separations are functions of the observed $\nu_{\rm max}$.

\begin{figure}
\centering
\includegraphics[angle= 90, width=12cm]{\fig/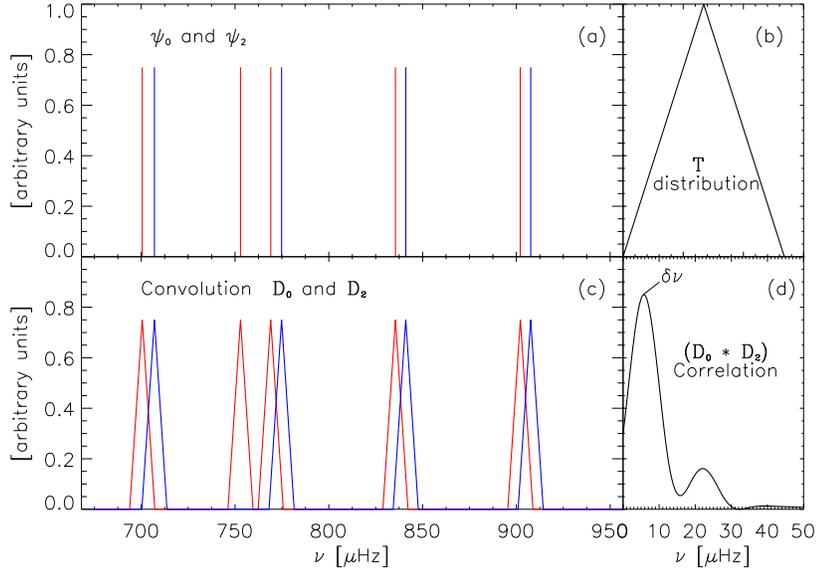}
\caption{ 
Description of the $\delta \nu$ computation. The functions $\psi_0, \psi_2$ corresponding to the computed frequencies for $l=0,2$, shown in blue and red respectively in panel (a), are convolved with the distribution $T$ in panel (b) to create the distribution $D_0,2$ in panel (c). The resulting cross-correlation function ($D_0 \ast D_2$) fixing the $\delta \nu$ at its maximum is shown in panel (d). The extra bump of the cross-correlation is caused by the extra $l=2$ mode at $\nu \sim 760$~$\mu$Hz and has no influence on the computed separation. 
\label{fig:correl}}
\end{figure}

The computation of the average large and small separations is done by finding the maximum of the cross-correlation function of the frequencies selected around a suitable value $\nu_{\rm central}$ near $\nu_{\rm max}$. Figure \ref{fig:correl} shows how the small separation is computed for a specific model of the grid. In panel (a), the function $\psi_0$ corresponding to the $l=0$ eigenfrequencies is shown in blue and the function $\psi_2$ for $l=2$ in red. These functions are convolved with the triangular distribution $T$ of panel (b). The base of the triangular distribution is one fifth of the smallest distance between any combination of the $l=0$ eigenfrequencies. The convolution
\begin{eqnarray}\label{convol}
D_{i}(x) \equiv \int_{-\infty}^{\infty} \psi_i(s) T(x-s) ds & i=0,2
\end{eqnarray} 
 is presented in panel (c) while the resulting cross-correlation, 
\begin{eqnarray}\label{cross}
(D_0 \ast D_2)(x) \equiv \int_{-\infty}^{\infty} D_0(s)D_2(x+s) ds &,
\end{eqnarray}
is in panel (d). The maximum of the correlation is at $\delta\nu$. It is easy to see that the extra $l=2$ mode around $\nu \sim 750$~$\mu$Hz is only producing a small bump in the correlation function and has no effect on the value of $\delta\nu$. For $\Delta \nu$, the location of the maximum of the auto-correlation $\left(D_0 \ast D_0\right)$, apart from the maximum at zero frequency shift, is taken. This technique is more robust than using the mean value of Equation (\ref{large}) and especially (\ref{small}) over different values of $n$.  The cross-correlation is not influenced by missing or extra eigenmodes, usually nonradial, produced by ADIPLS or by ambiguous determination of the radial order $n$ of a mode. The method is especially robust when it comes to computing  $\delta \nu$ for models having a convective core, or for models with avoided crossing. Robustness is very desirable since it is not possible to control the ADIPLS output for the 6 million models included in th
 e grid.

In the present version of SEEK, we computed all adiabatic modes with $l=0$, $2$ 
{ in a fixed range of the dimensionless frequency,} $10 \le \sigma^2 \le 2800$, where $\sigma$ is related to the angular mode frequency $\omega$ by
\begin{equation}\label{dimless}
\sigma^2 = \frac{R^3}{GM}\ \omega^2 \; , 
\end{equation}  
where $G$ is the gravitational constant. This range makes it possible to cover observed solar-like pulsation in known stars on the main sequence and subgiant branch. 
{ 
These pulsations cluster around the value of $\nu_{\rm max}$ which tends} to follow the scaling relation of \citet{KjeBed95}
\begin{equation}\label{scale}
\nu_{\rm max} \propto \frac{M}{R^2T^{1/2}}. 
\end{equation} 
{
To accelerate the SEEK procedure and for models showing eigenmodes in these region, we precalculated 25 different sets of $\Delta \nu$ and $\delta \nu$ for values of $\nu_{\rm central}$ equally spaced on a logarithmic scale, ranging from the Nyquist frequency of the short cadence of {\it Kepler}, 8330~$\mu{\rm Hz}$ (corresponding to a period of 2\,min) to 27.8~$\mu{\rm Hz}$ (corresponding to a period of 10\,h). 
(For any given model the actual range in $\nu_{\rm central}$ was restricted to
values corresponding to acoustic modes for that model.)}
If, for example, we want to compare the grid value of $\delta \nu$ with the solar value of $\delta \nu_{\odot}$ at $\nu_{\rm max,\odot} \sim 3333$~$\mu{\rm Hz}$, our grid automatically chooses the set $\delta \nu(\nu_{\rm central})$ which is the closest to $\nu_{\rm max, \odot}$ at $\nu_{\rm central} = 3220$~$\mu{\rm Hz}$.

\section{BAYESIAN APPROACH}

The Bayesian statistics method of SEEK and the notation used here were inspired mainly by the work of \citet{PonEye04} and \citet{JoerLin05}. These investigations compared the Bayesian approach to other means used for the determination of stellar age. We refer the reader to these papers and to the textbook by \citet{Gregory05} for technical questions involved in the use of Bayesian statistics.

Let us first define the maximum likelihood function 
\begin{equation}\label{likely}
\mathcal{L} = \left({\displaystyle\prod_{i=0}^n}\frac{1}{\sqrt{2\pi}\sigma_i}\right)\exp(-\chi^2/2),
\end{equation} 
with $\chi^2$ defined in Equation (\ref{chi2}). A maximum-likelihood estimate of stellar parameters can be obtained by finding the maximum of $\mathcal{L}$, which, in case of Gaussian errors, is equivalent to minimizing $\chi^2$. It can be argued that a maximum likelihood estimator is often enough to estimate stellar parameters. However, the highly non-linear mapping function $\mathcal{K}$ makes it necessary to provide a set of priors, especially if a realistic estimate of the errors on the stellar parameters is to be made. The prior can be seen as a specific weight attached to every point of the grid determining the probability of that point to be observed. The most obvious example of the utility of priors is well studied in \citet{JoerLin05}. It arises when one has to choose between different stellar models at different evolutionary stages but showing the same observables. In their example, one model is slowly evolving on the main sequence while the other is rushing up the 
 subgiant branch. An experienced astronomer would naturally choose the slowly evolving star as the most probable model even if the resulting $\mathcal{L}$ value is the same for both models. For the astronomer using Bayesian tools, a larger prior weight $f_0(\boldsymbol{p})$ given to the slowly evolving model automatically makes it a better candidate because the resulting probability, or posterior $f$ defined at each point $j$ as
\begin{equation}\label{baye}
f^j \propto f_0^j \mathcal{L}^j \; ,
\end{equation}
is bigger for that model. The grid overall posterior can be described as a density function,
\begin{equation}\label{baye_dense}
f(\boldsymbol{p}) \propto f_0(\boldsymbol{p}) \mathcal{L}(\mathcal{K}(\boldsymbol{p})) \; .
\end{equation}

>From the grid presented in the previous section, the computation of $\mathcal{L}$ at every point is straightforward. On the other hand, the definition of the priors $f$ requires some insight in the problem. 

The most natural way to write the prior density is as a function of $\boldsymbol{p}$, the model parameters, 
\begin{equation}\label{prior1}
f_0 = \Psi(\tau) \Phi(Z|\tau) \zeta(X_0|Z,\tau) \xi(M|X_0,Z,\tau) \beta(\alpha|X_0,Z,\tau,M)
\end{equation} 
where $\Psi(\tau)$ is the star formation rate (SFR) through time, and $\Phi(Z|\tau)$ and $\zeta(X_0|Z,\tau)$ describe the metallicity and initial hydrogen mixture as a function of age and as a function of age and metallicity, respectively. The initial mass function (IMF) $\xi(M|X_0,Z,\tau)$ is function of the element mixture as well as a function of time, and the prior related to the mixing-length parameter $\beta(\alpha|X_0,Z,\tau,m)$ can also depend on the other stellar parameters.

The distribution and the correlations can be built with the help of assumptions made from observations and from stellar and galactic models. In our case, since we are mainly interested in the {\it Kepler} satellite field of view which is well documented through the Kepler Input Catalog (KIC), we could use Equation (\ref{grid}) to map a color-color distribution, or any other relation found in the KIC, into the prior of Equation (\ref{prior1}). However, we shall keep here a more conservative approach, or as \citet{JoerLin05} put it, a non-committal approach and assume that $\tau$, $Z$, $X_0$, $M$ and $\alpha$ are independent. Thus we write
\begin{equation}\label{prior2}
f_0 = \Psi(\tau) \Phi(Z) \zeta(X_0) \xi(M) \beta(\alpha) \; ,
\end{equation} 
and also take $\Psi(\tau)$, $\Phi(Z)$, $\zeta(X_0)$, and $\beta(\alpha)$ to be flat. This means that the prior density is constant through these dimensions. We only make an assumption on the IMF and choose the so-called IMF1 model from \citet{Chabrier01} where
\begin{eqnarray}\label{IMF}
\xi(M) &=& \left\{ \begin{array}{ll} 0.019\ M^{-1.55} &\text{if } M \leq  1.0 \Msun \\
                                   0.019\ M^{-2.70} & \text{if } M >  1.0 \Msun .
                 \end{array}\right.
\end{eqnarray}

In a grid, flat priors can be easily illustrated by $N$ equally spaced points, every point having an equal probability $1/N$. Of course, if the grid is not regularly spaced the weight assignment can get a bit more complicated. For a non-flat prior, we could distribute the $N$ grid points so that they follow the density prescribed for example by Equation (\ref{IMF}) and give every point a weight of $1/N$ and meet the non-flat IMF prior prescription. However, and this is especially true in the $\tau$ direction, we have an irregularly spaced grid. Also, we do not wish to interpolate our grid for the sake of precision and for the same reason it is not necessarily reasonable to make the grid sparser in the region where the prior is relatively small. A star observed in that sparser region, even if less likely, would not be analyzed in the same detail by SEEK.

We have solved this problem by dividing the parameter space into a number of small 5-D bins of (penta-)volume $V^i$ containing each a small number of point $n^i \geq 1$. Then, a prior weight $f^i_0 $, where
\begin{equation}\label{prior3}
f^i_0 = \int_{V^i} \Psi(\tau) \Phi(Z) \zeta(X_0) \xi(M) \beta(\alpha) dV^i \; ,
\end{equation} 
is distributed through the $n^i$ point contained in the bin
such that the weight $f_{0k}^i$ given to model $k$ is inversely proportional to the evolution speed of that model,
\begin{eqnarray}\label{delta_tau}
f_{0k}^i = \frac{\Delta \tau_k }{\sum_{k'} \Delta \tau_{k'}} f_0^i& {\rm where} & k = 1,2,3,...,n^i \; .
\end{eqnarray}
Here $\Delta \tau_k$ is the time taken by model $k$ to evolve to the next point in its evolution track. This forces the slowly evolving models of the bin to have larger priors. These two equations are central pieces of SEEK. They ensure that the probability given to a model reflects the size of the parameter space where no other models are computed around it. These empty regions of the parameter space will be seen as gaps in some of the probability distributions of the next section.

\section{RESULTS}

In practice, once $f$ is fixed, the computation of the posterior probability $\mathcal{G}$ on an arbitrary set of parameters is easily done. In one dimension, the grid is sliced along the desired parameter and all models entering a bin from $s$ to $s+\Delta s$ are added 
\begin{equation}\label{probsum}
\mathcal{G}(s_i) = C {\displaystyle\sum_{k=0}^K} f^k(s_i) \; ,
\end{equation}
where the $f^k(s_i)$ are the $K$ points of the grid lying from $s_{i}$ to $s_{i+1} = s_i+\Delta s$, and $C$ is a constant ensuring that the total probability of the problem is 1 (see Equation \ref{probint}). The results of this computation is a probability distribution $\mathcal{G}(s)$ that is best seen as a histogram. The results for four parameters $R$, $M$, $\tau$, and $\Teff$, computed for 70~Ophiuchi~A, are presented in Figure \ref{fig:oph}. For convenience, there and in the following figures, we have normalized the histograms so that ${\rm max}(\mathcal{G}(s)) = 1$.

Another important quantity is the running integral of $\mathcal{G}$ seen as the solid red line in Figure \ref{fig:oph}. The running integral over the entire parameter space is the total probability of the grid. This should be 1 since we assume that the observed star lies within the grid. We use that relation to fix the constant $C$ in Equation (\ref{probsum})
\begin{equation}\label{probint}
{\displaystyle\sum_{i}} \mathcal{G}(s_i)= 1 \; .
\end{equation}
Similarly, we can integrate over a section of the distribution to get a $1$, $2$, or $3\,\sigma$ probability range for $s$. The error interval $[s^{(1)}, s^{(2)}]$ output by SEEK is by default at the $1\,\sigma$ level and is determined based on the probability integral of $\mathcal{G}$. Specifically,
\begin{equation}\label{errint}
\int_{s_0}^{s^{(1)}} \mathcal{G}(s) d s = 0.1585 \; , \qquad
\int_{s_0}^{s^{(2)}} \mathcal{G}(s) d s = 0.8415 \; ,
\end{equation}
where $s_0$ is the initial value of the parameter, and the integrals are represented by suitable sums, as in Equation (\ref{probint}). (Note that $0.8415- 0.1585 = 0.6830$ corresponds to $1\,\sigma$ for a normal distribution.) The solution SEEK outputs is the mid-point $(s^{(1)} + s^{(2)})/2$ of this interval. In this way the error bars are symmetric; this would not be the case if the median value $s^{(m)}$, with
\begin{equation}
\int_{s_0}^{s^{(m)}} \mathcal{G}(s) d s = 0.5 \; ,
\end{equation}
had been chosen. This $1\,\sigma$ section of the running integral is depicted in blue (between the dots) on the running integral of Figure \ref{fig:oph}.

These histograms along with the net numbers output by SEEK are useful to interpret the results. In the case of 70~Oph~A, the solutions $R = 0.86\pm   0.01$~[$\Rsun$] and  $\Teff = 5306\pm  36$~[K] are well constrained and quasi Gaussian. The histograms visually confirm the validity of the numerical solutions. The result for the mass $M = 0.90\pm   0.04$~$\Msun$ reveals some of the grid's limitation as there are gaps in the otherwise well defined structure of $\mathcal{G}(M)$. These gaps are artifacts of the 0.02~$\Msun$ mesh used in the grid below 1.8~$\Msun$. However, since the prior density is properly weighted through Equations (\ref{probsum}) and (\ref{probint}), the overall value of the mass, with its error bars, is correct. For the age, $\tau = 8.65\pm  3.64$~Gyr, a large error is revealed by the output and the histogram shows the probability distribution to be populated all along $\tau$ from 2~Gyr all the way to 13~Gyr. In that case a more reasonable answer is a lower 
 limit of $\tau \gtrsim 2$~Gyr. The histogram for $\tau$ also shows gaps that are due to the finite resolution of the grid.

Table \ref{tab:input} shows our selection of solar-like oscillators and observables found in the literature. These observables are processed by SEEK and produce the outputs in Table \ref{tab:compare} and Table \ref{tab:output}. In Table \ref{tab:compare}, we compared SEEK results with direct measurements of radius via interferometry and of mass via Kepler's third law. For SEEK, the fits have been made for two different posterior sets, the full set
\begin{equation}
\label{fullset}
\boldsymbol{q}^{\rm g,f} = 
\{\Delta\nu,\delta\nu,\Teff,\log g,{\rm [Fe/H]}, V, \pi, E_{B-V} \} \; ,
\end{equation}
and one using only a selected subset of the same observables,
\begin{equation}
\label{subset}
\boldsymbol{q}^{\rm g,s} = \{\Delta\nu,\Teff,{\rm [Fe/H]}\} \; .
\end{equation}
The subset may be more representative of what we can expect from a typical {\it Kepler} observation when no ground-based follow-up has been done for a star, and where the asteroseismic data have not yielded a reliable determination of $\delta \nu$. This can occur in a stochastically oscillating star when the width of observed modes is larger than the small separation itself. In that case, we have $\Delta \nu$ from the satellite observations, while $\Teff$ and [Fe/H] can be obtained from the KIC.

The results of Table \ref{tab:compare} show that the 3\% promise on the radius precision mentioned in the introduction is generally reached with both the full set and the subset of observables. More striking and certainly more important than the precision itself is the very good accuracy reached by SEEK. It is on this basis that our method, or any other technique, should be judged, provided that independent measurements are available. SEEK has pinned down all stellar radii at the 3\% level when the extended set of observable was used. The same level of accuracy is generally reached if the subset of observables (Eq.\ \ref{subset}) is used. 

{
We note that in well-posed problems, where parameters are actually strongly constrained, the value of $\mathcal{L}$ is much larger than $f_0$ close to the best solution. It is only when we get away from the best parameter values that the prior gets more important. The details of the priors used here only have an influence on the wing of the $\mathcal{G}$ distribution. Thus, the choice of priors has an effect on the errors, especially if the errors are large, but little effect on the solution itself. 
}

\section{DISCUSSION} 

\begin{figure}
\centering
\includegraphics[angle= 90, width=15cm]{\fig/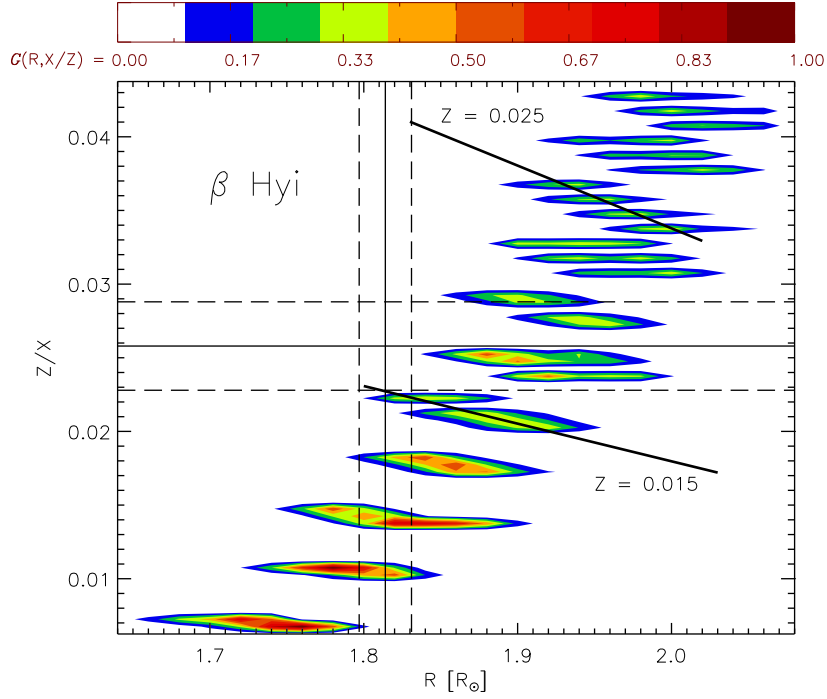}
\caption{ Correlation between $Z/X$ and $R$ for $\beta$~Hyi. The two thick solid curves through peaks in the correlation function $\mathcal{G}(R,Z/X)$ follow the constant metallicity ridges for $Z=0.025$ and 0.015. Along these curves the hydrogen mass ratio varies from the upper left peak to the lower right one as $X_0 = 0.68$, 0.70, 0.72, and 0.74.
A clear general trend is that $R$ increases with $Z$ and that it also increases with increasing $X_0$.
The solid horizontal line is the $\beta$~Hyi metallicity ratio of $Z/X = 0.0258 \pm 0.003 $ \citep{daS06},
with its error bars as dashed lines, while the solid and dashed vertical lines show the direct interferometric measurement of $R = 1.814\pm0.107$. Note that the merging of the constant metallicity ridges at lower values is an artifact of the drawing procedures.
\label{fig:cor_beta}}
\end{figure}

We clearly need to understand the exceptions to the general success of SEEK; in particular, we note that $\beta$~Hyi was a bit further from the direct observations with an offset of 6\%. An extra tool offered by SEEK can be used to study the $\beta$~Hyi radius offset when compared to direct measurements. Figure \ref{fig:cor_beta} shows a two-dimensional projection $\mathcal{G}(R,Z/X)$ of the posterior computed for $\beta$~Hyi with only $\Delta \nu$ and $\Teff$ as input observables. The metallicity is not included as an observable in the computation as a trick to have a clearer view of the correlation between that observable and $R$. Several possibilities are offered to explain the discrepancy between SEEK's radius $R/R_\odot =1.92 \pm0.05$ and the direct measurement of $1.814\pm 0.017$. The first possibility is that the observed $Z/X$ is too high. Reducing [Fe/H] by $ \sim 0.2$ dex, to $Z/X \sim 0.0163$, would reconcile the asteroseismic and interferometric radii. This can be
  concluded from a visual examination of Figure \ref{fig:cor_beta}. In fact, lowering  the metallicity by this amount would put it back to the level measured in $\beta$~Hyi by \citet{DraLinVan98}. We have made an a posteriori test using the value of metallicity and temperature published by \citet{DraLinVan98}, $\Teff = 5800 \pm 100$\,K, and [Fe/H]$ = 0.2 \pm 0.1 $ instead of the more recent values of Table \ref{tab:input} \citep{daS06}. These inputs processed by SEEK place the radius $R/R_\odot  = 1.85 \pm 0.03$ in much better agreement (2\%) with the observations. We note that \citet{DraLinVan98} used the older \citet{NoeGre93} metal mixtures in the computation. This underlines that accuracy can only be reached if the input metallicity is selected with some care. 

The offset in radius for $\beta$~Hyi can be explained otherwise. The extension of the grid to lower values of $X_0$ would to some extent reconcile SEEK and the interferometric measurement. It can be seen in Figure \ref{fig:cor_beta} that the left-hand side of the correlation function $\mathcal{G}$ is in fact the edge of the grid where $X_0=0.68$. The extension of the grid would likely expand the existing ridge of solutions to the left in Figure \ref{fig:cor_beta}, at lower radii. A simple eye-ball estimate in the figure can be made to indicate that the combination, $Z \sim 0.0165$, $X_0 \sim 0.64$ would yield the right metallicity and the right radius. However, it seems that a hydrogen content that low is not reasonable and perhaps this possibility should not be considered seriously. The effect of the hydrogen content, or should we say helium content, on the radius is in fact the main factor compromising the precision of $R$ in SEEK or any other technique using the large sepa
 ration to determine the radii. The error on the radius is influenced by the size of the grid in the $X_0$ direction. The imperative we had not to underestimate the errors on stellar parameters influenced our choice of having a large range of values for that parameter. It also means that raising the precision on the metallicity ratio $Z/X$ to fix the radius with better precision has some intrinsic limitations, unless the helium abundance can be fixed independently. We also note that the gaps between the islands of Figure~\ref{fig:cor_beta}, like the one for the 1D histograms, are caused by the finite resolution of the grid. The gaps in this figure also reflect that if the grid is regularly spaced in $X_0$ and $Z$ it is not in $Z/X$ since the gaps are not regularly spaced along that direction.

One other result obtained by SEEK is worth examining in more detail. For Procyon~A, if the subset of observables is used to do the fit, the values of $M$ and $R$ are less precise, which is in accordance with expectation, but more accurate than the case where all observables are used. This could reveal some inconsistency in the observables as well as some of the limitations of the SEEK procedure. Indeed forcing SEEK to look for a model of Procyon~A with the right astrometry, $V=0.363$ and $\pi = 134.07$\,mas, reduces the accuracy of the results, compared with using only $\Teff$, [Fe/H] and $\Delta \nu$. This discrepancy could come from inhomogeneity between the tool used in the spectroscopic study giving the observables and in our models. The metallicity and temperature taken as input for Procyon~A are from 3D modeling \citep{All02}, while we used the conversion of the \citet{VanCle03} tables using 1D models to get our value of $M_V$. Using older atmospheric parameters, or a f
 ully consistent conversion, could perhaps improve the fit. 

We also remark that the behavior of our results for the mass of $\alpha$~Cen~B presents similarity to those of Procyon~A. The results are slightly more accurate if $V$, $\pi$ and $\delta \nu$ are dropped from the fit. It seems that this reveals some of the limitation of our grid when extremely precise measurements are available for a star. The small box defined within $\sim 1-\sigma$ of all $\alpha$~Cen~B observables only includes a few models. This number is too small and can only produce a weak probabilistic assessment over the value of the mass. Computing more models to cover with more precision the parameters space of $\alpha$~Cen~B would most likely resolve the small deviation over the value of the mass. This underscores the fact that modeling dedicated to individual stars becomes useful once classical observables of a star are very precise.

In any case, Table~\ref{tab:compare} shows that SEEK is certainly capable of pinning down the radius of solar-like oscillators with a minimum effort and that it is also able to fix the mass of these stars with an accuracy not attainable when only non asteroseismic inputs are available. This improvement in accuracy is especially potent in case the distance to the star is not known.

\begin{figure}
\begin{center} 
$\begin{array}{c}
\includegraphics[angle= 0, width=12cm]{\fig/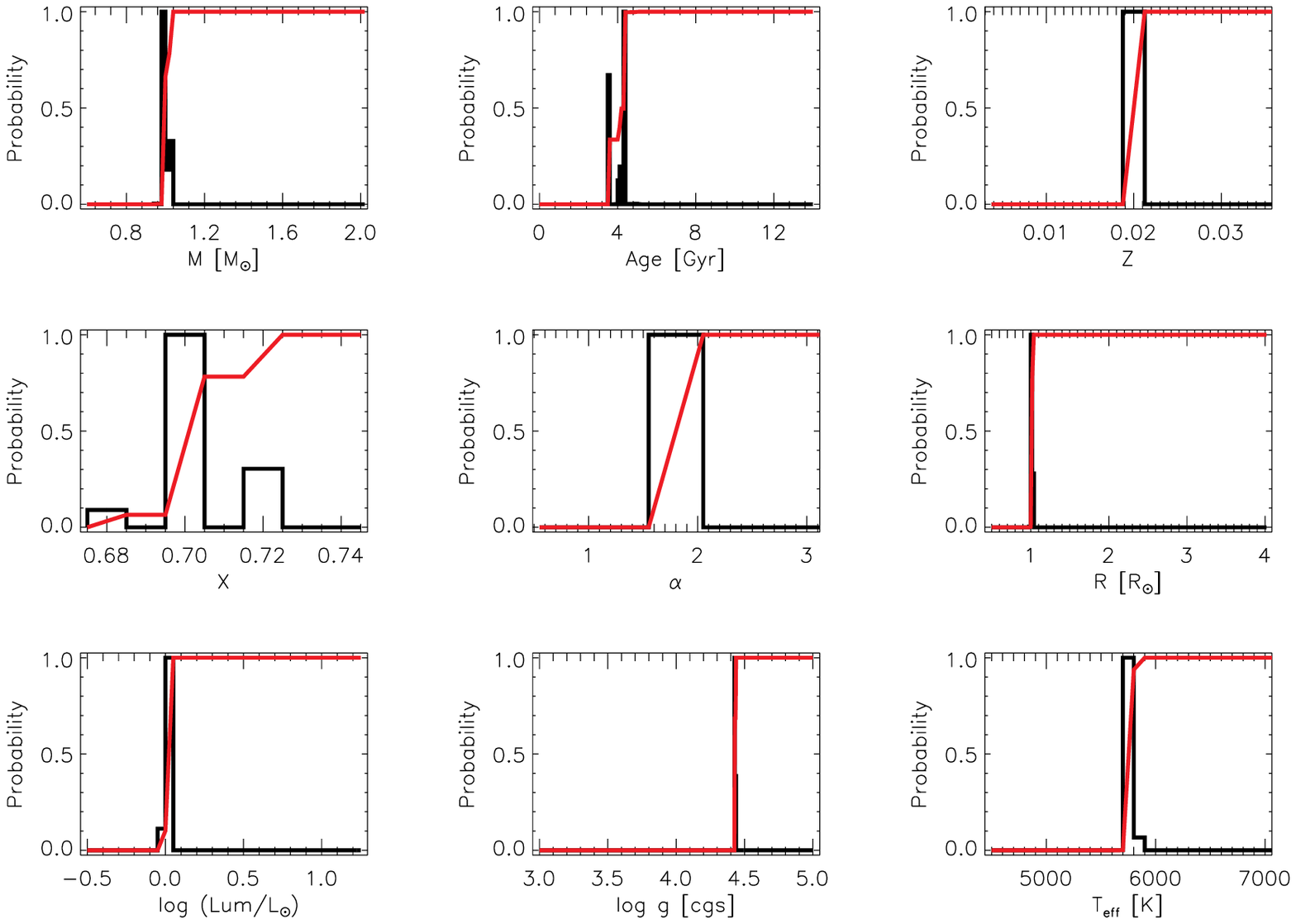}
\\
\\
\\
\includegraphics[angle= 0, width=12cm]{\fig/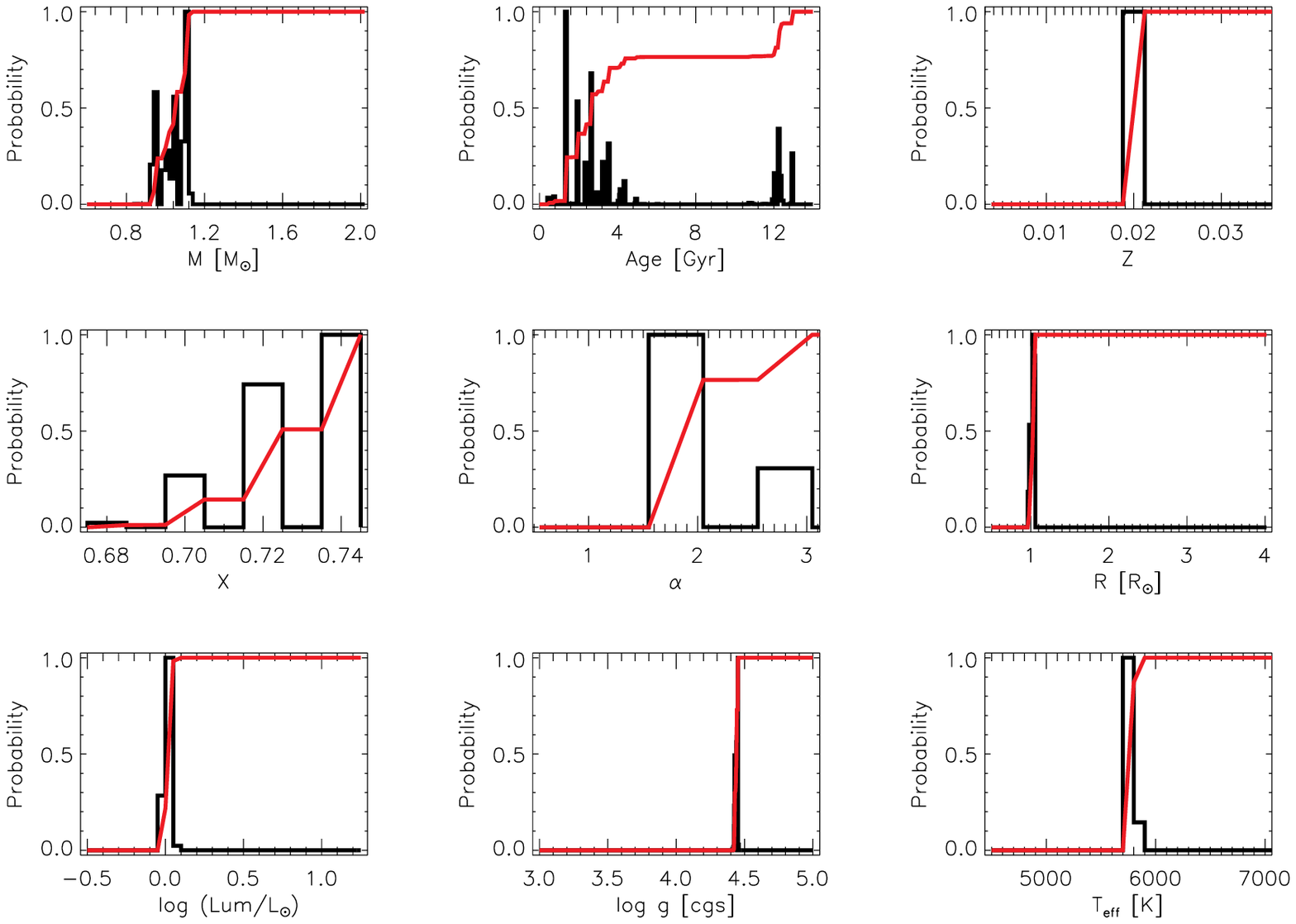}
\end{array} $
\end{center}
\caption{SEEK's posterior for the Sun. The top 9 panels represent the probability distribution (normalized to 1) as well as the probability integral for the model using all the observables of Table \ref{tab:input}. The lower panels are similar but use a subset of observables ($\Delta \nu$, $\Teff$, and [Fe/H]).   
\label{fig:sun}}
\end{figure}

In Table~\ref{tab:output}, all primary stellar parameters $\boldsymbol{p}$ of the modeled stars are shown as well as some selected secondary values. We address here the question of the age of stars fixed by SEEK, and especially the age of the Sun, since it is the only star where an independent determination of the age is available. A fairly good accuracy, 3.96$\pm$0.41~Gyr (13\%), compared to the meteoritic value of 4.57$\pm$0.02~Gyr \citep[Wasserburg, in][]{BahPinWas95}, is reached in the case where $\Delta \nu$ and $\delta \nu$ are known. A look at the histogram in the top panel of Figure~\ref{fig:sun} confirms visually that the solution is well constrained. This is quite promising in the case where the small separation is known. It seems that we can expect a precision of 5 to 20\% for the age when $\Delta \nu$, $\Teff$, [Fe/H], and especially $\delta \nu$ are known with reasonable precision. 

If only the three observables in the subset are used, the result varies from star to star.  For the Sun, the lower panel of Figure~\ref{fig:sun} shows for the age two distinct and very probable solutions. This specific example shows how useful the histograms are since the output of Table~\ref{tab:output} does not reveal the presence of two solutions. One thing not shown here, but that we were able to see from the computation of the probability distribution  $\mathcal{G}(\tau,\alpha)$, is that the two islands of solution for the age are at two different value of $\alpha$, the young solution corresponding to the standard value of $\alpha = 1.8$ and the older one to $\alpha =  2.8$. As seen in the lower panel of Figure \ref{fig:sun} higher values of $\alpha$ are acceptable if only the subset of observables is used.

\begin{figure}
\begin{center} 
$\begin{array}{c}
\includegraphics[angle= 0, width=12cm]{\fig/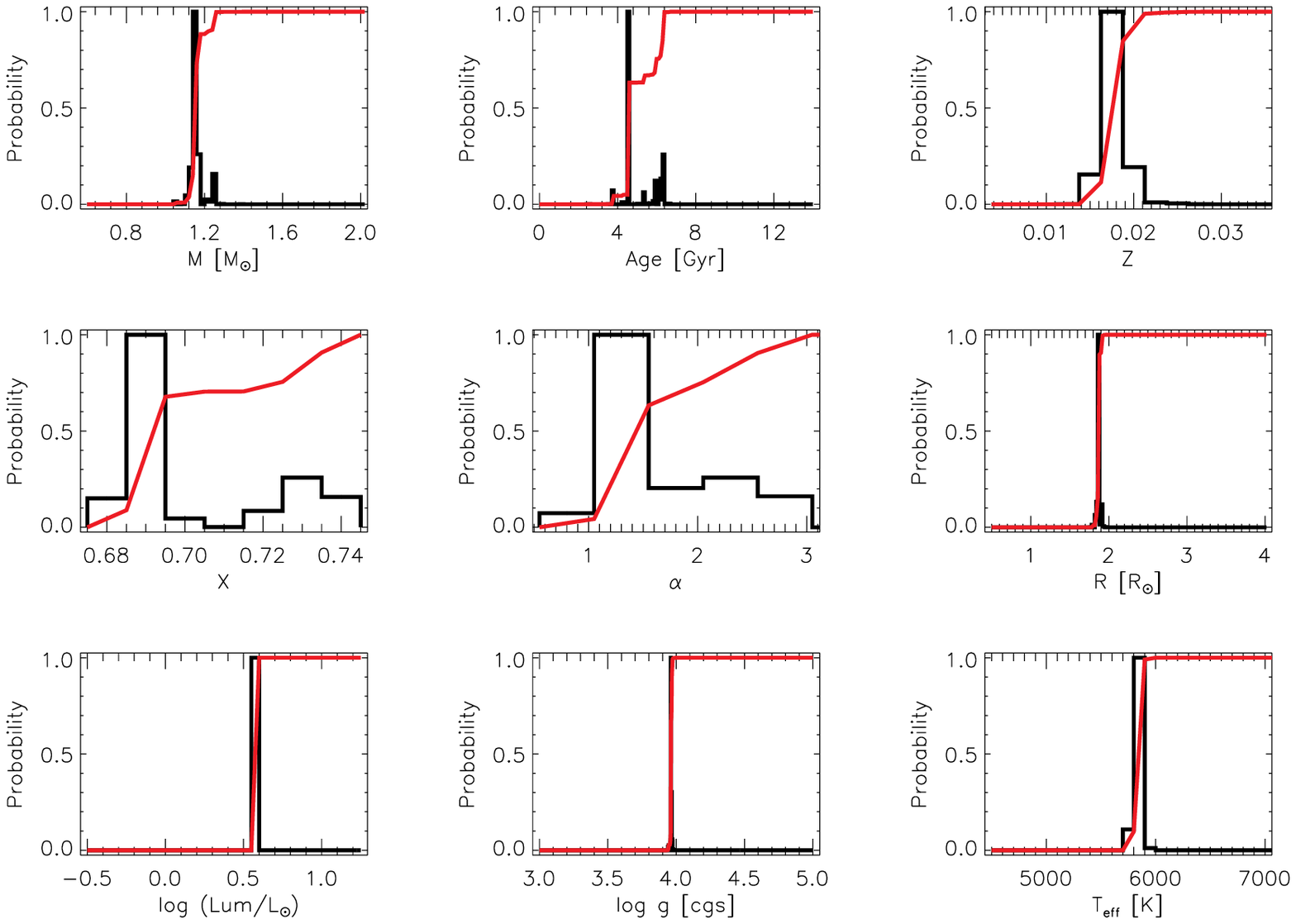}
\\
\\
\\
\includegraphics[angle= 0, width=12cm]{\fig/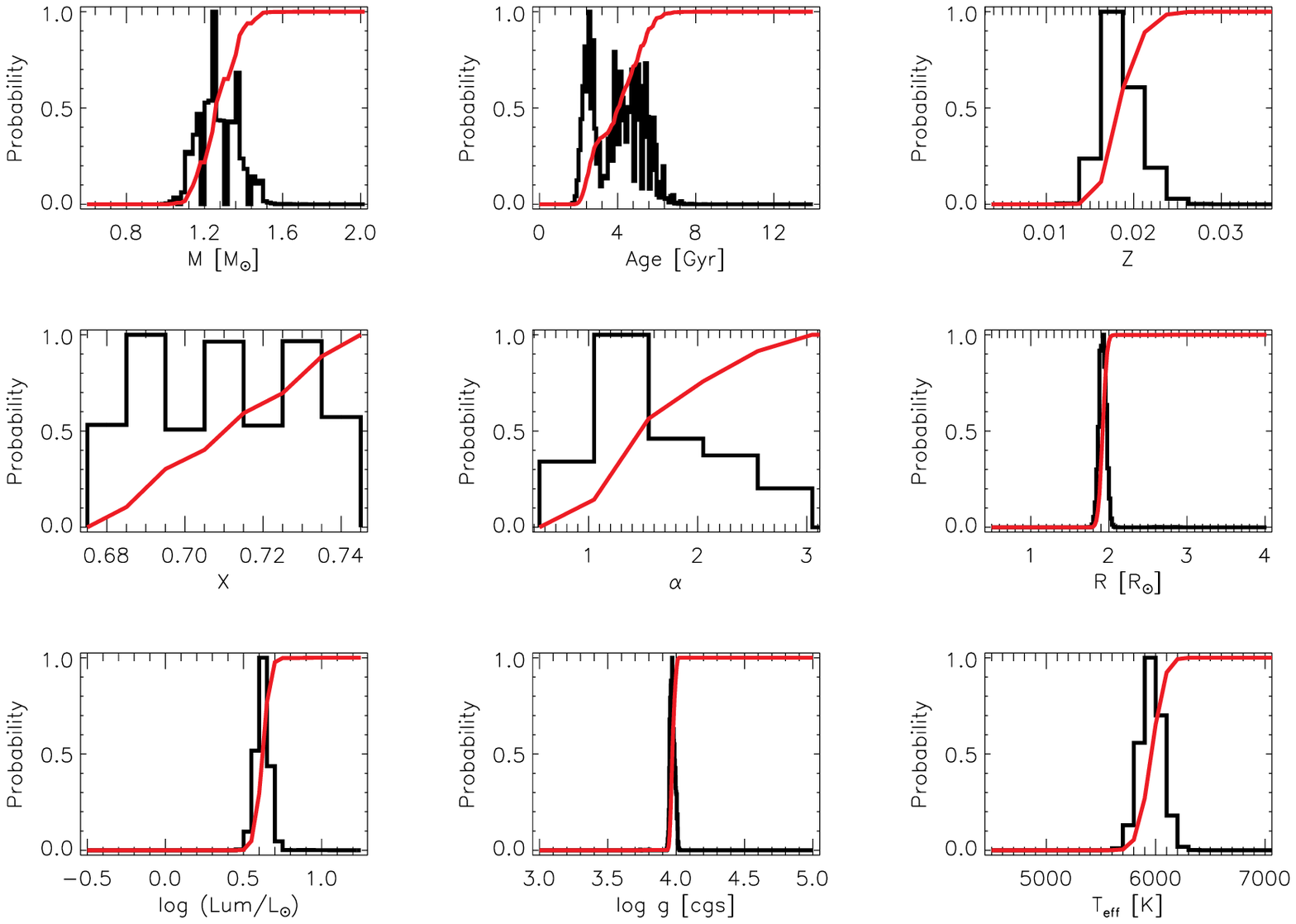}
\end{array}$
\end{center}
\caption{Same as Figure \ref{fig:sun} for $\beta$~Hyi. \label{fig:beta}}
\end{figure}

Figure \ref{fig:beta} illustrates another type of effect on the age when the subset of observables is used. The precision on the age of $\beta$~Hyi is degraded by 20\%, but the overall answer, or accuracy, is the same. As mentioned previously in the description of Figure \ref{fig:oph}, 70~Oph~A shows a third type of scenario where not much can be concluded on the age of the star as the posterior is showing solutions at almost any age. Since the relation between the age and the observables is highly non-linear, it is very difficult to estimate beforehand what will be the precision obtained on the age depending on the type of star and/or on the precision of the observables, especially when $\delta \nu$ is not known. 

We also note that any age determination based on stellar modeling is susceptible to changes in basic aspects of the modeling. In the solar case helioseismic fits to low-degree modes result in a solar age very close to the meteoritic value, for models including diffusion and settling of helium and heavy elements, and when the `old' solar composition is used, whereas use of the \citet{Asplun2009} composition induces a significant shift \citep[see][for a review]{Christ2009}. Also, the age determination based on fits to the {\it Kepler} data for HAT-P-7, which has a convective core, was understandably rather sensitive to the extent to which convective-core overshoot was included \citep{Christ2010}. Such potential systematic effects must clearly be taken into account in the interpretation of results of fits such as those carried out with SEEK. Of course the long-term goal is to reduce these effects through an improved understanding of stellar structure and evolution based on more 
 detailed asteroseismic investigations.

\section{CONCLUSIONS}

We have shown the details involved in the SEEK procedure and we have tested the validity of the approach with all independent measurement known to us. This is the first step before {\it Kepler} observations of solar-like stars, now becoming available in large quantity \citep{Gillil2010}, are analyzed. We hope that SEEK will be able to process these observation with ease and lead to the publication of a catalog that includes a homogeneous sample of stars revealing their radius, mass, age, and other stellar parameters.  

{
We expect that the results of SEEK can be a useful starting point in more detailed asteroseismic analyses in the, likely frequent, cases where extensive sets of oscillation frequencies are available. Such detailed investigations can further refine the basic stellar parameters, including the age which, as shown above, is sometimes not significantly constrained by SEEK.
Also, detailed analyses of the frequencies are likely to uncover evidence for the need for improvements in stellar modeling.}

{
The SEEK model grid extends to the base of the red-giant branch and hence
SEEK can be used to determine properties of stars on both the main sequence and
the subgiant branch. 
In fact, the results obtained for the subgiants $\eta$ Boo and $\beta$ Hyi
indicate the potential of SEEK for the investigation of subgiants,
common amongst the {\it Kepler} asteroseismic targets.
On the other hand we acknowledge the substantially different diagnostic 
potential between centrally hydrogen burning stars and subgiants.
In particular, $\delta \nu$ provides a direct measure of stellar age
in the former case,
while the diagnostic potential of $\delta \nu$ for subgiants is more subtle.
In the latter case, detailed analysis of individual frequencies of mixed
modes may provide much more stringent constraints \citep[e.g.,][]{Metcal2010}.
These issues of optimizing the asteroseismic diagnostics certainly
require further investigations.}

In parallel, further tests and development of SEEK are required. A first step will be to test the level of systematic errors that are introduced by the neglect of relevant physical effects in the stellar modeling. An obvious example is diffusion and settling which have been demonstrated to have a substantial effect on solar modeling, as tested with helioseismology. Also, effects of convective core overshoot could influence the results for stars slightly more massive than the Sun. New grids of models including such effects can then be computed for inclusion in SEEK.

A related issue, with less obvious solutions, concerns the systematic effects of the frequency errors introduced by the failure to model properly the outermost layers of the star, which are known to dominate the difference between observed and modeled frequencies in the solar case \citep[e.g.,][]{Christ1996} and which also have potentially significant effects on the large and small separations. It is important at least to obtain an estimate of the extent to which these effects may influence the results obtained with SEEK. One may hope that improvements in the modeling of the relevant effects, combined with detailed analyses of {\it Kepler} data for a broad range of stars, will eventually allow us to reduce or eliminate such potential systematic errors.

\acknowledgments

POQ would like to thank H. Bruntt, J. Rowe, G. Do{\u g}an, T. Campante and J. Dupuis for their interest in this work. This project was supported by the European Helio- and Asteroseismology Network (HELAS), a major international collaboration funded by the European Commission's Sixth Framework Programme, by the Danish Natural Science Research Council and by the Natural Sciences and Engineering Research Council of Canada through a Canadian Space Agency Visiting Fellow grant.

\bibliographystyle{apj}
\bibliography{apj-jour,bibliographie}

\begin{figure}
\centering
\includegraphics[angle= 90, width=12cm]{\fig/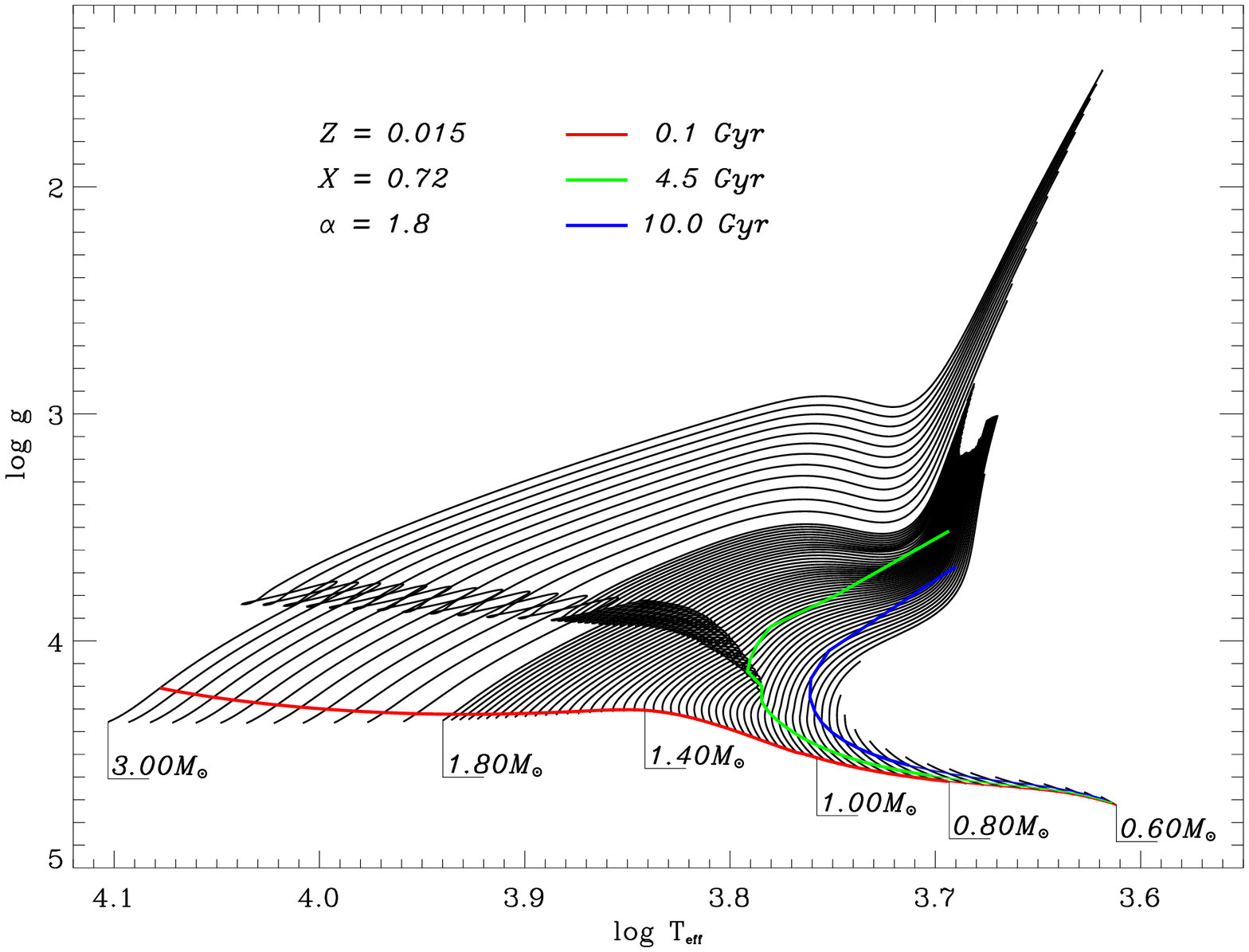}
\caption{ Slice of the SEEK grid projected in the $\loggteff$ diagram for 
$Z =0.015$ $X_0 = 0.72$ and $\alpha=1.8$. Three isochrones are also drawn. The computation was stopped after the star had reached 15 billion years or an arbitrary point on the RGB branch. The low-mass models are separated by steps of 0.02~$\Msun$, the more massive ones by 0.10~$\Msun$. 
\label{fig:slice}}
\end{figure}

\begin{figure}
\centering
\includegraphics[angle= 90, width=12cm]{\fig/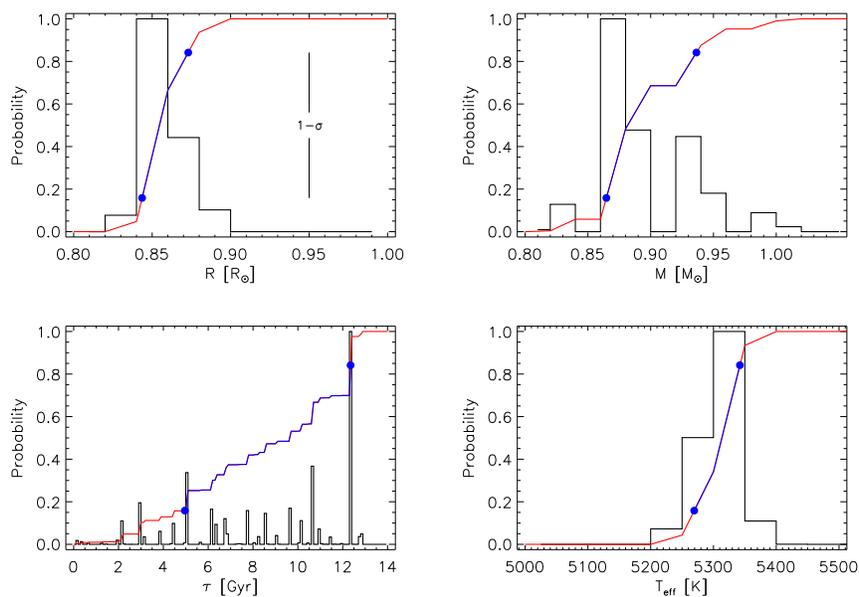}
\caption{SEEK's posterior for 70 Oph A. All the observables of Table \ref{tab:input} are used in the fit. The probability distribution $\mathcal{G}(p)$ for $R$, $M$, $\tau$ and $\Teff$, (normalized to a maximum of 1, black histogram) is plotted along with the probability integral (solid red line). The delimitation of the $1\,\sigma$ probability of the distribution, determined by the interval $[s^{(1)}, s^{(2)}]$ (cf.\ Eq. \ref{errint}), is plotted between the blue dots.
\label{fig:oph}}
\end{figure}

\begin{deluxetable}{rccccccccccc}
\tabletypesize{\tiny}
\rotate
\tablecolumns{12}
\tiny
\tablewidth{0pt}
\tablecaption{Observables
\label{tab:input}}
\tablehead{  
star && $\Delta \nu$ $[\mu \text{Hz}]$ & $\delta \nu$ $[\mu \text{Hz}]$& $\nu_{\rm max}$ $[\mu \text{Hz}]$&$\Teff$ [K] & [Fe/H]& $\log g$ [cgs] & $V$ & $\pi$ [mas] &$E_{B-V}$& Reference

}
\startdata
$\beta$ Hyi       && $57.2\pm0.5$&$5.3\pm0.5$ &$1000$ &$5964\pm 100$ &$-0.03\pm0.05$&$-$&$2.8\pm0.01$&$134.07\pm0.11$& $ 0.01$ & 1 ;  a \\
$\tau$ Cet        && $169.6\pm0.5$ &$12.7\pm1.2$ &$4500$ &$5264\pm100$ &$-0.5\pm0.03$&$4.36\pm0.15 $ &$3.39\pm0.01$ &$-$ &$0$& 2 ; b \\
Procyon A         && $55.5\pm0.5$ &$-$ &$1000$ &$6514\pm27$  &$-0.05\pm0.05$  &$-$  &$0.363\pm0.003$ &$285.93\pm0.88$ & $0$ & 3 ; c\\
$\eta$ Boo      && $39.9\pm0.9$ &$4.0\pm1.0$ &$750$ &$6007\pm255$  &$0.09\pm0.01$ &$-$  &$2.68\pm0.03$ &$88.17\pm0.75$ &$0$& 4 ; d\\
$\alpha$ Cen A    && $105.5\pm0.5$ &$5.6\pm0.7$ &$2410$ &$5847\pm27$  &$0.24\pm0.03$ &$4.34\pm0.12$  &$-0.01\pm0.01$ &$747.1\pm1.2$ &$0$& 5 ; e \\
$\alpha$ Cen B    && $161.5\pm0.5$ &$10.1\pm0.6$ &$4100$ &$5316\pm28$  &$0.25\pm0.04$  &$ 4.44\pm0.15$  &$1.35\pm0.01$ &$747.1\pm1.2$ &$0$& 6 ; e \\
70 Oph A          && $161.7\pm0.8$ &$-$ &$4500$ &$5300\pm50$  &$0.04\pm0.05$  &$-$  &$4.19\pm0.014$ &$194.2\pm1.2$ &$0$& 7 ; f  \\
Sun               && $134.8\pm0.5$ &$9.1\pm0.2$ &$3034$ &$5778\pm20$ &$0\pm0.01$  &$4.44\pm0.01$    &$-$ &$-$ &$-$& 8 ; g \\
\enddata           
\tablerefs{Asteroseismology (1) \citealt{Bed07}; (2) \citealt{Tei09}; (3) \citealt{EggCarBou05}; (4)\citealt{CarEggBou05}; (5)\citealt{BouCar02}; (6) \citealt{Kje05}; (7) \citealt{CarEgg06}; (8) \citealt{Thi00} }
\tablerefs{Other observables (a) \citealt{daS06}; (b) \citealt{SouKatCay98}; (c) \citealt{All02}; (d) \citealt{Mor02}; (e) \citealt{PorLyrKel08}; (f) \citealt{Egg08} ; (g) \citealt{GreSau98}}
\end{deluxetable}

\begin{deluxetable}{rccccccccccc}
\tabletypesize{\tiny}
\rotate
\tablecolumns{12}
\tiny
\tablewidth{0pt}
\tablecaption{Comparison With Direct Measurements\label{tab:compare}}
\tablehead{      && \multicolumn{4}{c}{$R$ $[\Rsun]$} && \multicolumn{4}{c}{$M$ $[\Msun]$}& \\
                      \cline{3-6}                \cline{8-11} \\
                 && \multicolumn{2}{c}{SEEK}& & && \multicolumn{2}{c}{SEEK} \\
         Star    && Selected          & All & Direct/Interferometry& Reference  && Selected & All& Direct/Kepler's Law& Reference}
\startdata
$\beta$ Hyi    &&  $1.92\pm0.05$&$1.87\pm0.01$ & $1.814\pm 0.017$&1 &&$1.27\pm0.11$ &$1.16\pm0.02$  &                    &    \\
$\tau$ Cet     &&  $0.79\pm0.01$&$0.81\pm0.01$ & $0.790\pm 0.005$&2 && $0.77\pm0.03$&$0.83\pm0.01$  &                    &    \\
Procyon A      &&  $2.04\pm0.05$&$2.12\pm0.02$ & $2.048\pm 0.025$&3 &&$1.45\pm0.10 $&$1.59\pm0.05$  & $1.497 \pm 0.037$  &4   \\
$\eta$ Boo   &&  $2.65\pm0.08$&$2.69\pm0.06$ & $2.672\pm 0.028$&5 && $1.62\pm0.13$&$1.71\pm0.08$  &                    &    \\
$\alpha$ Cen A &&  $1.23\pm0.04$&$1.25\pm0.01$ & $1.224\pm 0.003$&6 && $1.09\pm0.09$&$1.11\pm0.01$  & $1.105 \pm 0.007$  &6   \\
$\alpha$ Cen B &&  $0.87\pm0.01$&$0.87\pm0.01$ & $0.863\pm 0.005$&6 && $0.92\pm0.04$&$0.89\pm0.01$  & $0.935 \pm 0.006$  &6   \\
70 Oph A       &&  $0.86\pm0.02$&$0.86\pm0.01$ &                 &  && $0.89\pm0.06$&$0.89\pm0.05$  & $0.890 \pm 0.020$  &7   \\
Sun            &&  $1.02\pm0.03$&$1.01\pm0.01$ & $1$             &  && $1.03\pm0.08$&$1.01\pm0.02$  & $1$                &    
\enddata           
\tablerefs{(1)\citealt{Nor07}; (2)\citealt{diF07}; (3)\citealt{Ker04}; (4)\citealt{Gir00} ;(5)\citealt{VanCiaBod07}; (6)\citealt{MigMon05}; (7)\citealt{Egg08};  }
\end{deluxetable}

\begin{deluxetable}{rlcccccccccccc}
\tablewidth{0pt}
\tabletypesize{\tiny}
\rotate
\tablecolumns{14}
\tiny
\tablecaption{SEEK Output \label{tab:output}}
\tablehead{  & &&  \multicolumn{11}{c}{Stellar Parameters} 
\\
	      &&   \multicolumn{5}{c}{Primary} && \multicolumn{5}{c}{Secondary} 
\\
		     \cline{4-8} \cline{10-14} \\
	      Star & Observables && $M$ $[\Msun]$ &$\tau$ [Gyr]  &$Z$ &$X_0$ &$\alpha$
		  && $R $ $[\Rsun] $ &$\log (L/\Lsun)$ &$\log g$ [cgs]  &$\Teff$ [K] &$Y$ }
\startdata
$\beta$ Hyi& All  && $1.16\pm0.02 $&$5.41\pm0.89 $&$0.018\pm0.001 $&$ 0.708\pm0.022 $&$ 1.74\pm0.60 $    
		  && $1.87\pm0.01 $&$0.58\pm0.02 $&$3.964\pm0.004 $&$ 5845\pm38   $&$ 0.274\pm0.023$  
\\& Selected
                  && $1.27\pm0.11 $&$3.94\pm1.45 $&$0.019\pm0.002 $&$ 0.710\pm0.022  $&$1.69\pm0.62 $
		  && $1.92\pm0.05 $&$0.62\pm0.05 $&$3.97\pm0.02  $&$ 5959\pm111  $&$0.271\pm0.024 $ 
\\
\\
$\tau$ Cet&All	  && $0.83\pm0.01 $&$0.75\pm0.03 $&$0.007\pm0.001 $&$ 0.730\pm0.003$&$ 1.30\pm0.17 $	
		  && $0.81\pm0.01 $&$-0.32\pm0.02 $&$4.547\pm0.006 $&$ 5450\pm34   $&$ 0.263\pm0.004$ 
\\& Selected

    		  && $0.77\pm0.03 $&$6.31\pm2.18 $&$0.007\pm0.001 $&$ 0.729\pm0.004  $&$1.30\pm0.17$
    		  && $0.79\pm0.01 $&$-0.36\pm0.03 $&$4.53\pm0.01  $&$ 5286\pm71    $&$0.264\pm0.005$ 
\\
\\
Procyon A&All	  && $1.59\pm0.05 $&$1.83\pm0.24 $&$0.021\pm0.003 $&$ 0.731\pm0.012 $&$ 2.42\pm0.46$	
		  && $2.12\pm0.02 $&$0.88\pm0.02 $&$3.991\pm0.010  $&$ 6548\pm37   $&$ 0.248\pm0.015$ 
\\& Selected
                  && $1.45\pm0.10 $&$2.25\pm0.64 $&$0.018\pm0.002 $&$ 0.710\pm0.023  $&$1.93\pm0.62$
      	          && $2.04\pm0.05 $&$0.82\pm0.02 $&$3.98\pm0.02  $&$ 6513\pm64    $&$0.272\pm0.025$ 
\\
\\
$\eta$ Boo&All	  && $1.71\pm0.08 $&$1.90\pm0.38 $&$0.025\pm0.001 $&$ 0.729\pm0.013 $&$ 1.32\pm0.60$    
		  && $2.69\pm0.06 $&$0.97\pm0.02 $&$3.81\pm0.02  $&$ 6166\pm80   $&$ 0.246\pm0.014$ 
\\& Selected
      	          && $1.62\pm0.13$ &$2.24\pm0.62$ &$0.025\pm0.001$ &$ 0.728\pm0.013$ &$1.33\pm0.64$
      	          && $2.65\pm0.08$ &$0.88\pm0.08$ &$3.80\pm0.02$  &$ 5974\pm280$ &$0.247\pm0.014$ 
\\
\\
$\alpha$ Cen A& All   && $1.11\pm0.01 $&$4.95\pm0.04 $&$0.030\pm0.001 $&$ 0.680\pm0.003$&$ 1.80\pm0.17$	
		  && $1.25\pm0.01 $&$0.18\pm0.02 $&$4.299\pm0.001 $&$ 5850\pm34   $&$ 0.290\pm0.004$ 
\\& Selected
      	          && $1.09\pm0.09$ &$8.04\pm4.05$ &$0.030\pm0.001$ &$ 0.698\pm0.020$ &$2.25\pm0.59$
      	          && $1.23\pm0.04$ &$0.20\pm0.03$ &$4.30\pm0.01$  &$ 5850\pm37$ &$0.272\pm0.021$ 
\\
\\
$\alpha$ Cen B &All   && $0.89\pm0.01 $&$6.95\pm0.03 $&$0.030\pm0.001 $&$ 0.680\pm0.003 $&$ 1.80\pm0.17$
		  && $0.87\pm0.01 $&$-0.32\pm0.02 $&$4.521\pm0.001 $&$ 5250\pm34   $&$ 0.290\pm0.004$ 
\\& Selected
   
		  && $0.92\pm0.04$ &$6.99\pm4.45$ &$0.029\pm0.001$&$ 0.709\pm0.030$ &$2.23\pm0.52$
		  && $0.87\pm0.01$ &$-0.26\pm0.03$ &$4.53\pm0.01$&$ 5309\pm66$ &$0.262\pm0.020$ 
\\
\\
70 Oph A&All	  && $0.89\pm0.05 $&$8.65\pm3.70 $&$0.023\pm0.003 $&$ 0.716\pm0.018 $&$ 2.06\pm0.34$	
		  && $0.86\pm0.01 $&$-0.27\pm0.02 $&$4.525\pm0.006  $&$ 5311\pm65   $&$ 0.261\pm0.021 $ 
\\& Selected
 
      	          && $0.89\pm0.06$&$ 7.54\pm4.54$&$ 0.022\pm0.003 $&$ 0.710\pm0.026$&$ 1.99\pm0.33$
      	          && $0.86\pm0.02$&$-0.28\pm0.03$&$ 4.52\pm0.01  $&$ 5304\pm71$ &$0.268\pm0.029$ 
\\
\\
Sun&	All	  && $1.01\pm0.02 $&$3.96\pm0.41 $&$0.020\pm0.001 $&$ 0.707\pm0.011 $&$ 1.80\pm0.17 $	
		  && $1.01\pm0.01 $&$0.02\pm0.02 $&$4.432\pm0.005 $&$ 5753\pm36   $&$ 0.273\pm0.012 $ 
\\& Selected
      	          && $1.03\pm0.08$&$ 6.80\pm5.43$&$ 0.020\pm0.001 $&$ 0.729\pm0.013$&$ 2.18\pm0.53$
      	          && $1.02\pm0.03$&$ 0.03\pm0.03$&$ 4.44\pm0.01  $&$ 5757\pm39$ &$0.251\pm0.014$
\\

\enddata	   
\end{deluxetable}

\end{document}